\documentclass[12pt,english]{revtex4}
\usepackage{amsmath}
\usepackage[T1]{fontenc}
\usepackage[latin9]{inputenc}
\usepackage{bm}
\usepackage{epsfig}
\usepackage{babel}
\usepackage{babel}
\usepackage{babel}

\makeatletter
\@ifundefined{textcolor}{}
{
 \definecolor{BLACK}{gray}{0}
 \definecolor{WHITE}{gray}{1}
 \definecolor{RED}{rgb}{1,0,0}
 \definecolor{GREEN}{rgb}{0,1,0}
 \definecolor{BLUE}{rgb}{0,0,1}
 \definecolor{CYAN}{cmyk}{1,0,0,0}
 \definecolor{MAGENTA}{cmyk}{0,1,0,0}
 \definecolor{YELLOW}{cmyk}{0,0,1,0}
 }
\@ifundefined{textcolor}{}{ \definecolor{BLACK}{gray}{0}
 \definecolor{WHITE}{gray}{1}
 \definecolor{RED}{rgb}{1,0,0}
 \definecolor{GREEN}{rgb}{0,1,0}
 \definecolor{BLUE}{rgb}{0,0,1}
 \definecolor{CYAN}{cmyk}{1,0,0,0}
 \definecolor{MAGENTA}{cmyk}{0,1,0,0}
 \definecolor{YELLOW}{cmyk}{0,0,1,0}
 }\@ifundefined{textcolor}{}{ \definecolor{BLACK}{gray}{0}
 \definecolor{WHITE}{gray}{1}
 \definecolor{RED}{rgb}{1,0,0}
 \definecolor{GREEN}{rgb}{0,1,0}
 \definecolor{BLUE}{rgb}{0,0,1}
 \definecolor{CYAN}{cmyk}{1,0,0,0}
 \definecolor{MAGENTA}{cmyk}{0,1,0,0}
 \definecolor{YELLOW}{cmyk}{0,0,1,0}
 }
\makeatother
\input{tcilatex}
\begin{document}

\title{Dynamics of diffusion controlled chain closure: flexible chain in
presence of hydrodynamic interaction }
\author{Rajarshi Chakrabarti}

\begin{abstract}
Based on the Wilemski-Fixman approach (J. Chem. Phys. 60, 866 (1974)) we showed that for a flexible chain in $%
\theta$ solvent hydrodynamic interaction treated with an pre-averaging
approximation makes ring closing faster if the chain is not very short. Only
for a very short chain the ring closing is slower with hydrodynamic
interaction on. We have also shown that the ring closing time for a chain with
hydrodynamic interaction in $\theta$ solvent scales with the chain length ($%
N $) as $N^{1.527}$, in good agreement with previous renormalization group calculation based prediction by Freidman et al. (Phys. Rev. A. 40, 5950 (1989)).
\end{abstract}

\address{Department of Inorganic and Physical Chemistry, Indian Institute
of Science, Bangalore 560012}

\maketitle

\section{introduction}

Dynamics of loop formation in long chain molecules has been a subject of immense interest to
experimentalists \cite{winnikbook, haung, lapidusjpcb, hudgins} and
theoreticians \cite{wilemski, doi, perico, szabo, pastor, sebastianjcp,
cherayiljcp20021, cherayiljcp20022, cherayiljcp2004, portman, sokolov, toan, santo}
for over a decade. Recently the dynamics of loop formation has found greater
importance because of its relevance in biophysics. Advances in single
molecule techniques made it possible to monitor the kinetics of loop
formation involving biomolecules at the single molecule level \cite%
{Cloutier, thirumalai}. Loop formation is a prime step in protein \cite%
{eaton} and RNA folding \cite{thirumalai2001}. Also a measure of the intrinsic
flexibility of DNA can be found from the rate at which it undergoes
cyclization \cite{Cloutier}.

Loop formation in polymers is essentially a many body problem as a polymer is made of many connected segments and hence an exact
analytical solution is impossible. All the theories of loop closing dynamics
are approximate \cite{wilemski, szabo}. Although many theoretical
and simulation attempts have been made to investigate the effect of
flexibility \cite{santo}, solvent quality \cite{cherayiljcp2004} on the loop
closing dynamics, not much theoretical investigation has
been performed to shed light on the effects of hydrodynamic
interaction on loop closing dynamics other than the renormalization group calculation by Freidman et al \cite{friedman}.
Hydrodynamic interaction which is
essentially nonlocal in space has been shown to have profound effects on the
dynamics of long chain molecules \cite{sainpre, pablo}. The rate of
translocation of polymer through a nano-pore has been shown to be greatly
affected in presence of hydrodynamic interaction \cite{pablo, ali}. Recently it
has been theoretically shown \cite{sainpre} that the breakage rate of
stretched polymer tethered to soft bond get enhanced in presence of
hydrodynamic interaction.

In this paper we investigate the effect of hydrodynamic interaction on the ring
closing dynamics of a flexible chain in $\theta$ solvent. It is well known
that the ring closing time ($\tau$) for a flexible chain without excluded volume and hydrodynamic
interaction (Rouse chain
\cite{doibook}) scales with the chain length ($N$) as $N^{2}$ in the
Wilemski-Fixman (WF) \cite{wilemski} theoretical framework. Here we use Zimm model \cite%
{doibook} for the polymer which actually takes care of the hydrodynamic
interaction at the simplest possible level. Our calculation shows that
hydrodynamic interaction profoundly affect the rate of loop formation and
also has a different scaling relation with the length of the polymer $N$.
The rest of the paper is arranged as follows. In Sec. II the Wilmeski-Fixman (WF)
theory for the chain closure \cite{wilemski} is briefly discussed. WF
formalism gives a prescription to calculate the ring closing time as an
integral over a sink-sink time correlation function. Sec. III deals with a
brief description of the radial delta function sink which is used later in
the calculation of the ring closing time. Time correlation formalism for the
flexible chain with and without hydrodynamic interaction is discussed in
Sec. IV. Sec. V presents the results and VI is devoted to conclusions.

\section{Theory of Chain Closure}

In an stochastic environment the dynamics of a single polymer chain having
reactive end-groups is modeled by the following Smoluchowski equation in WF
theory \cite{wilemski}.

\begin{equation}
\frac{\partial P(\{\mathbf{R}\},t)}{\partial t}=LP(\{\mathbf{R}\},t)-kS(\{%
\mathbf{R}\})P(\{\mathbf{R}\},t)  \label{Smolu1}
\end{equation}

Here $P(\{R\},t)$ is the distribution function for the chain that it has the
conformation $\{R\}\equiv R_{1},$$R_{2},$$......R_{n}$ at time $t$ where $%
R_{i}$ denotes the position of the $i$th monomer in the chain of $n$
monomers. $S({R})$ is called the sink function which actually models the
reaction between the ends and thus usually is a function of end to end
vector. $L$ is a differential operator, defined as

\begin{equation}
L=D_{0}\sum_{i=1}^{n}\frac{\partial }{\partial \mathbf{R}_{i}}.\left[ \frac{%
\partial }{\partial \mathbf{R}_{i}}+\frac{\partial U}{\partial \mathbf{R}_{i}%
}\right] P(\{\mathbf{R}\},t)  \label{operator}
\end{equation}

Here $D_0$ is the diffusion coefficient of the chain defines as the inverse
of the friction coefficient per unit length and $U$ is the potential energy
of the chain. Wilemski and Fixman then derived an approximate expression for
the mean first passage time from Eq. (\ref{Smolu1}). This mean first passage
time is actually the loop closing time for the chain. The expression for
this loop closing time reads

\begin{equation}
\tau =\int_{0}^{\infty }dt\left( \frac{C(t)}{C(\infty )}-1\right)
\label{tau}
\end{equation}

\noindent Where $C(t)$ is the sink-sink correlation function defined
as

\begin{equation}
C(t)=\int d\mathbf{R}\int d\mathbf{R}_{0}S(\mathbf{R})G(\mathbf{R},t|\mathbf{%
R}_{0},0)S(\mathbf{R}_{0})P(\mathbf{R}_{0})  \label{ct}
\end{equation}

\noindent In the above expression $G(\mathbf{R},t|\mathbf{R}_{0},0)$ is the Greens
function or the conditional probability that a chain with end-to-end
distance $R_0$ at time $t=0$ has the end-to-end distance $R$ at time $t$; $%
P(R_0)$ is the equilibrium distribution of the end to end distance since the
chain was in equilibrium at time $t=0$. $S(R)$ is the sink function \cite%
{sebastianpra1992, debnath, chakrabarticpl2010} which depends only on the
separation between the chain ends. We would like to comment that Eq. (\ref%
{tau}) is not exact and only valid in the limit of infinite sink strength, $%
k\to\infty$ as this closing time is nothing but the mean first passage time \cite{RednerBook, chakrabartijcp}.

Now it is obvious that the knowledge of the Greens function, $G(\mathbf{R}%
,t| \mathbf{R}_{0},0)$ is prerequisite to calculate the sink-sink
correlation function $C(t)$ and hence the closing time $\tau$. In case of a
flexible chain the Greens function and the end-to-end probability
distribution functions are known and Gaussian.

For a flexible chain the Greens function is given by

\begin{equation}
G(\mathbf{R},t|\mathbf{R}_{0},0)=\left( \frac{3}{2\pi \left\langle
R^{2}\right\rangle _{eq}}\right) ^{3/2}\frac{1}{\left( 1-\phi ^{2}(t)\right)
^{3/2}}\times \exp \left[ -\frac{3(\mathbf{R}-\phi (t)\mathbf{R}_{0})^{2}}{%
2\left\langle R^{2}\right\rangle _{eq}(1-\phi ^{2}(t))}\right]
\label{greens}
\end{equation}

Where

\begin{equation}
\phi (t)=\frac{\left\langle \mathbf{R}(t).\mathbf{R}(0)\right\rangle _{eq}}{%
\left\langle R^{2}\right\rangle _{eq}}  \label{phi}
\end{equation}

is the normalized end-to-end vector correlation function for the chain. The
above ensemble average is taken over the initial equilibrium distribution
for end-to-end vector $P(\mathbf{R}_{0})$.

\bigskip

Similarly end-to-end equilibrium distribution for the flexible chain at time
$t=0$ is given by \cite{doibook, sokolov}

\begin{equation}
P(R_{0})=\left( \frac{3}{2\pi L^{2}}\right) ^{3/2}\exp \left[ -\frac{%
3R_{0}^{2}}{2L^{2}}\right]  \label{peq}
\end{equation}

With the above Gaussian functions the sink-sink correlation function can be
written as a radial double integral.

\begin{eqnarray}
C(t) &=&\left( \frac{3}{2\pi L^{2}}\right) ^{3}\frac{1}{\left( 1-\phi
^{2}(t)\right) ^{3/2}}\int_{0}^{\infty }4\pi R^{2}S(R)dR\int_{0}^{\infty
}4\pi R_{0}^{2}S(R_{0})dR_{0}\times  \notag \\
&&\exp \left[ -\frac{3}{2L^{2}}\frac{(R^{2}+R_{0}^{2})}{(1-\phi ^{2}(t))}%
\right] \frac{\sinh \left[ (3\phi (t)RR_{0})/(L^{2}(1-\phi ^{2}(t)))\right]
}{(3\phi (t)RR_{0})/(L^{2}(1-\phi ^{2}(t)))}  \label{sink-sink}
\end{eqnarray}

The above integral can be evaluated analytically for some specific choice of
the sink functions. With a radial delta function sink the above integral can
be evaluated analytically.

\section{The radial delta function sink}

For our case study we choose a radial delta function sink \cite{pastor}, $S(R)=\delta
(R-a) $. Since in this case the integration over $R$ and $R_0$ can be
carried out analytically, the looping time can be expressed in a closed form
as follows.

\begin{equation}
\tau =\int_{0}^{\infty }dt\left( \frac{\exp \left[ -(2x_{0}\phi
^{2}(t))/(1-\phi ^{2}(t))\right] \sinh \left[ (2x_{0}\phi (t))/(1-\phi
^{2}(t))\right] }{2x_{0}\phi (t)\sqrt{1-\phi ^{2}(t)}}-1\right)
\label{taudelta}
\end{equation}

with
\begin{equation*}
x_0=\frac{3 a^2}{2 L^2}
\end{equation*}

Obviously if the end-to-end vector correlation function $\phi(t)$ is known
the closing time can be calculated by carrying out the integration over
time. Here we calculate the looping time given by the above
expression for a flexible polymer in presence and absence of hydrodynamic
interaction and make a comparative study. Throughout the paper it is assumed that the chain was in equilibrium at $t=0$ and only at $t=0^{+}$ the hydrodynamic interaction is turned on (for the Zimm chain). Thus the equilibrium end-to-end distribution for the Rouse as well as for the Zimm chain  is given by (Eq.(\ref{peq})). The last integration over time is
not analytical so has to be carried out numerically.

\section{Flexible polymer without and with hydrodynamic interactions}

The simplest dynamical description of a flexible chain or polymer in
solution is given by Rouse Model \cite{doibook, kawakatsubook}. This model
does not take into account of excluded volume and hydrodynamic interaction
and has been the basis of dynamics of dilute polymer solution and has been
successfully used in the context of ring closing in polymers \cite%
{sebastianjcp}, breathing dynamics in DNA \cite{chakrabarti}, polymer
translocation \cite{klspre} etc. In the continuum limit the governing
equation of motion for the position vector $\mathbf{R}(n,t)$ is given by

\begin{equation}
\zeta \frac{\partial \mathbf{R}(n,t)}{\partial t}=k\frac{\partial ^{2}%
\mathbf{R}(n,t)}{\partial n^{2}}+f(n,t)  \label{rouse}
\end{equation}

\noindent Here $n$ is a continuous variable, $k=\frac{3 k_B T}{b^2}$ where $b$ is the
length of a monomer in the discrete representation of the Rouse polymer and $f(n,t)$ is the random force with the moments

\bigskip

\begin{eqnarray}
\left\langle f(n,t)\right\rangle &=&0  \notag \\
\left\langle f_{\alpha }(n,t_{1})f_{\beta }(m,t_{2})\right\rangle &=&2\zeta
k_{B}T\delta (n-m)\delta _{\alpha \beta }\delta (t_{1}-t_{2})
\label{random force}
\end{eqnarray}

It is straightforward to show that the Rouse normal modes ($X_{p}$) obey the
following equation

\bigskip

\begin{equation}
\zeta_p \frac{\partial {X}_{p}}{\partial t}=-k_{p}X_{p}+f_{p}(t)
\label{normalrouse}
\end{equation}

\noindent Where $\zeta_0=N\zeta$, $\zeta_p=2N\zeta$ for $p=0,1,2,...$
and $k_p=2\pi^2 k p^2/N=\frac{6 \pi^2 k_B T}{N b^2}p^2$ for $p=0,1,2,...$. Here $f_p(t)$ is the random force satisfying
$\left<f_{p \alpha}\right>=0$ and $\left<f_{p\alpha}(t)f_{q \beta}(t_1)\right>=2\delta_{pq}\delta_{\alpha\beta} \zeta_{p} k_B T\delta(t-t_1)$.

Then for a Rouse polymer it can be shown that the normalized end-to-end vector
time correlation function defined in (\ref{phi}) is given by

\begin{equation}
\phi(t)=\frac{\left\langle \mathbf{R}(t).\mathbf{R}(0)\right\rangle _{eq}}{%
\left\langle R^{2}\right\rangle _{eq}}=\sum_{p=odd}\frac{8}{p^{2}\pi ^{2}}%
e^{-tp^{2}/\tau _{1}}= \sum_{p=odd}\frac{8}{p^{2}\pi ^{2}}%
e^{-t\omega^{Rouse}_{p}}  \label{phi_rouse}
\end{equation}

\noindent Where $\tau_1=\frac{\zeta N^2 b^2}{3 \pi^2 k_B T}$ and $\omega^{Rouse}_{p}=%
\frac{p^2}{\tau _{1}}$.

The simplest possible model for the dynamics of a flexible chain with the hydrodynamic interaction but without
the excluded volume interaction  is
known as the Zimm model \cite{doibook, kawakatsubook}. In $\theta$ condition the
continuum limit equation of motion for the position vector $\mathbf{R}(n,t)$
in the Zimm model is given by

\begin{equation}
\zeta \frac{\partial \mathbf{R}(n,t)}{\partial t}=\sum_{m}H_{mn}.\left( k%
\frac{\partial ^{2}\mathbf{R}(m,t)}{\partial m^{2}}+f(m,t)\right)
\label{zimm}
\end{equation}

Here $H_{mn}$ is the mobility matrix and is generally a nonlinear function of $%
\mathbf{R}(n,t)-\mathbf{R}(m,t)$ and the above equation is quite difficult
to handle. To simplify this analysis Zimm introduced a pre-averaging
approximation, which replaces $H_{mn}$ by its equilibrium average, $%
\left\langle H_{mn}\right\rangle _{eq}$. \ It can be shown that with this
pre-averaging approximation \cite{doibook} becomes a linear function of $%
\mathbf{R}(n,t)$.

\begin{equation}
\zeta \frac{\partial \mathbf{R}(n,t)}{\partial t}=\sum_{m}h(n-m)\left( k%
\frac{\partial ^{2}\mathbf{R}(m,t)}{\partial m^{2}}+f(m,t)\right)
\label{zimm2}
\end{equation}

\noindent where $h(n-m)$ decreases slowly as $h(n-m)\propto \left|(n-m)\right|^{-1/2}$%
. Thus in Zimm model interaction among the segments is not localized. This is how a Zimm chain in different from a Rouse chain.

It is possible to show \cite{doibook}  that for the Zimm chain in $\theta$ solvent, the normal modes ($X^{Zimm, \theta}_{p}$) can be shown to obey an equation which has the same structure as the Rouse normal modes

\bigskip

\begin{equation}
\zeta_p \frac{\partial {X}^{Zimm, \theta}_{p}}{\partial t}=-k_{p}{X}^{Zimm, \theta}_{p}+f_{p}(t)
\label{normalrouse}
\end{equation}

\noindent Where $\zeta_p=\zeta\sqrt{\frac{\pi N p}{3}}$ for $p=0,1,2,...$
and $k_p=2\pi^2 k p^2/N=\frac{6 \pi^2 k_B T}{N b^2}p^2$ for $p=0,1,2,...$. Here $f_p(t)$ is the random force satisfying
$\left<f_{p \alpha}\right>=0$ and $\left<f_{p\alpha}(t)f_{q \beta}(t_1)\right>=2\delta_{pq}\delta_{\alpha\beta} \zeta_{p} k_B T\delta(t-t_1)$.

In case of $\theta$ solvent condition the normalized end-to end vector time
correlation function (Eq.(\ref{phi})) for the Zimm chain is given by

\begin{equation}
\phi(t)=\frac{\left\langle \mathbf{R}(t).\mathbf{R}(0)\right\rangle _{eq}}{%
\left\langle R^{2}\right\rangle _{eq}}=\sum_{p=odd}\frac{8}{p^{2}\pi ^{2}}%
e^{-tp^{3/2}/\tau^{Zimm,\theta} _{1}}= \sum_{p=odd}\frac{8}{p^{2}\pi ^{2}}%
e^{-t \omega^{Zimm, \theta}_{p}}  \label{phi_zimm}
\end{equation}

\noindent Where $\tau^{Zimm, \theta}_{1}=\frac{\zeta N^{3/2} b^2}{6 \sqrt{3} \pi^{3/2}
k_B T}$ and $\omega^{Zimm, \theta}_{p}=\frac{p^{3/2}}{\tau^{Zimm, \theta}_{1}%
}$.

In principle it is straightforward to calculate the closing time with a
delta function sink (Eq.(\ref{taudelta})) as it involves evaluating $\phi(t)$ and
carrying out an integration over time. We evaluate $\phi(t)$ for both the
cases exactly by carrying out the sums defined in (Eq.(\ref{phi_rouse})) and (Eq.(\ref{phi_zimm})).

\begin{figure}[tbp]
\centering
\epsfig{file=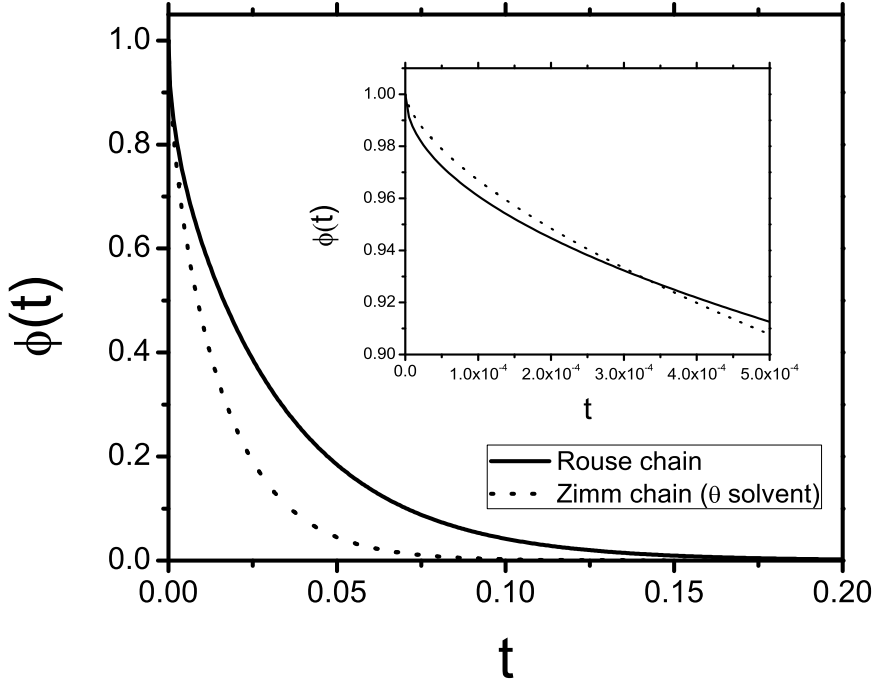,width=0.8\linewidth}\newline
\caption{$\protect\phi(t)$ against time ($t$). The values of other
parameters used are: $N=1, b=1, \protect\zeta=1, k_BT=1 $. In the inset the
short time ($t$) behavior of $\protect\tau$ is shown. Note up to a short
time range $\protect\phi(t)$ for the chain with hydrodynamic interaction
(Zimm chain in $\protect\theta$ solvent) decays slowly as compared to a
chain without hydrodynamic interaction (Rouse chain).}
\label{phitshort}
\end{figure}

\begin{figure}[tbp]
\centering
\epsfig{file=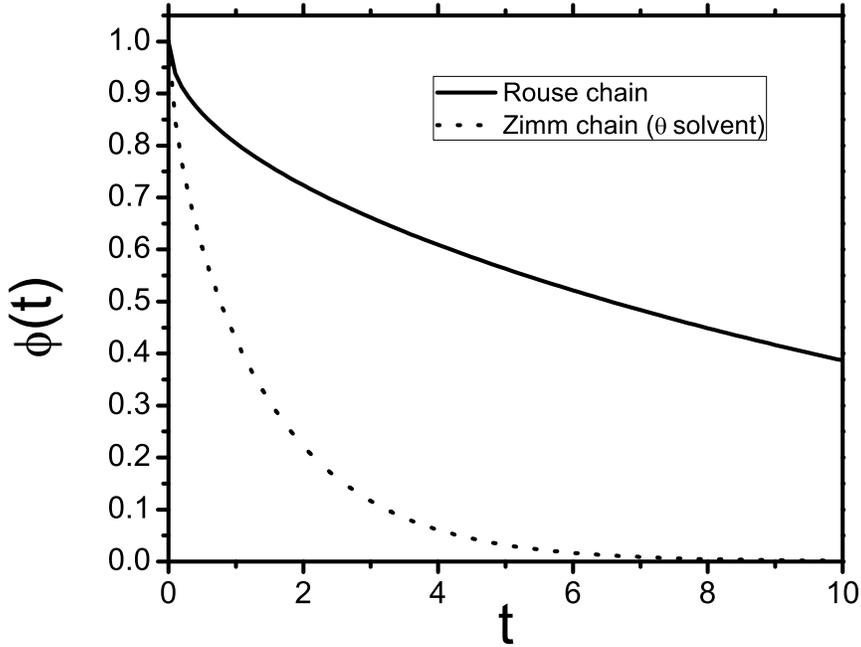,width=0.8\linewidth}\newline
\caption{$\protect\phi(t)$ against time ($t$). The values of other
parameters used are: $N=20, b=1, \protect\zeta=1, k_BT=1 $.}
\label{phitlong}
\end{figure}

\begin{figure}[tbp]
\centering
\epsfig{file=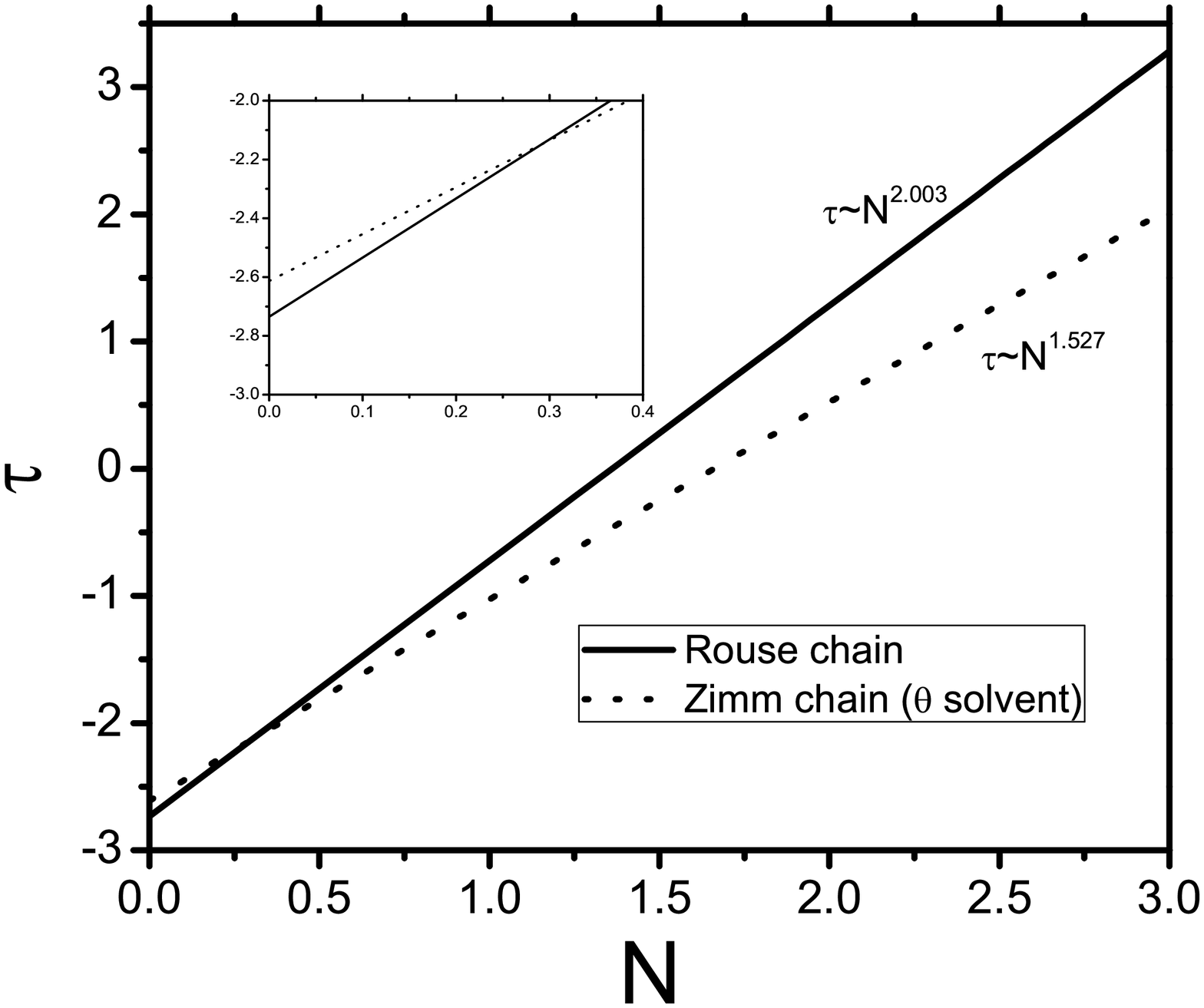,width=0.8\linewidth}\newline
\caption{$\protect\tau$ against the chain length($N$). The values of other
parameters used are: $b=1, \protect\zeta=1, k_BT=1, a=0.01 $. In the inset $%
\protect\tau$ for a short range of the chain length, $N$ shown. Note for a
very short chain the $\protect\tau$ for the chain with hydrodynamic
interaction (Zimm chain in $\protect\theta$ solvent) is higher meaning a
slower ring closure.}
\label{tau}
\end{figure}

\section{Results}

The closing time ($\tau$) with the radial delta function sink is calculated
for the Rouse chain as well as for the Zimm chain in $\theta$ solvent.
Calculation of $\tau$ involves evaluating $\phi(t)$ for the respective
chains and then putting it in Eq. (\ref{taudelta}) followed by an integration
over time which has to be carried out numerically. The end-to-end vector correlation function defined in Eq. (\ref{phi}) is
actually a sum over the decays of all the odd modes and at a given time $t$
the bulk of the contribution to $\phi(t)$ comes from the lower normal modes. Each normal mode decays with an
effective rate constant, $\omega^{Rouse}_{p}=p^2/\tau_{1}$ for the Rouse
chain (Eq.(\ref{phi_rouse})) and in case of a Zimm chain in $\theta$ solvent it is $\omega^{Zimm,
\theta}_{p}=p^{3/2}/\tau^{Zimm,\theta} _{1}$ (Eq.(\ref{phi_zimm})). Now had it been $%
\tau_{1}=\tau^{Zimm, \theta}_{1}$ all the odd modes of the Rouse chain hence
the end-to-end vector correlation function would have decayed faster
resulting a faster ring closure compared to that for a Zimm chain in $\theta$
solvent. In reality $\tau_{1}\neq\tau^{Zimm, \theta}_{1}$ and the ratio $%
\omega^{Zimm, \theta}_{p}/\omega^{Rouse}_{p}$ has a strong length ($N$)
dependence, $\omega^{Zimm, \theta}_{p}/\omega^{Rouse}_{p}=\frac{%
(\tau_1/\tau^{Zimm,\theta} _{1})}{\sqrt{p}}=\frac{2\sqrt{3}}{\sqrt{\pi}}
\sqrt{N/p}$. Only when, $\sqrt{N/p}<<\frac{2\sqrt{3}}{\sqrt{\pi}}$ the
normal modes of the Zimm chain in $\theta$ solvent decay faster. For a very
short chain or in case of higher normal modes the condition $\sqrt{N/p}<<%
\frac{2\sqrt{3}}{\sqrt{\pi}}$ can be achieved and then only the normal modes
for the Zimm chain (in $\theta$ solvent) will decay slowly compared to the
normal modes of the Rouse chain. Thus for a short Zimm chain (in $\theta$ solvent) initial decay of $\phi(t)$ is
slow compared to a Rouse chain. This is shown in the inset of Fig. 1. On the other hand for a long Zimm chain all the normal modes
decay faster than the Rouse chain normal modes. This naturally results a slowly decaying $\phi(t)$ for a Rouse chain as is shown in Fig. 2.

In Fig. 3 log-log plot of the closing time ($\tau$) is plotted against the chain length $N$ for both the chains. The slope for the Rouse chain has been found out to be $2.003$ very close to previously found values based on WF theory calculation \cite{cherayiljcp2004, thirumalai}. For the Zimm chain (in $\theta$ solvent) the slope is $1.527$. Thus hydrodynamic interaction makes ring closure faster. Also notice for a very short chain the ring closure is faster for a Rouse chain as shown in the inset of Fig. 3. This actually suggests that the proportionality factor or the frequency factor between the ring closing rate, $\tau$ and $N^{\alpha}$, where $\alpha=1.527$ with the hydrodynamic interaction on and $\alpha=2.003$ without the hydrodynamic interaction, too has a length ($N$) dependence.

\section{Conclusions}

In this paper we show that the hydrodynamic interaction makes ring closure faster within WF \cite{wilemski} framework if the chain is not very short. For a very short chain ring closure can be slower in presence of hydrodynamic interaction. The ring closing time ($\tau$) for the Zimm chain (in $\theta$ solvent) scales as $N^{1.527}$. This is actually in very good agreement with the renormalization group calculation prediction of Friedman et al. \cite{friedman}. They showed that for a Zimm chain not too short the closing time scales as $N^{3/2}$. But surprisingly $N^{3/2}$ behavior can also be seen even without hydrodynamic interaction. For example, harmonic chain approximation within WF framework leads to similar scaling \cite{doi}.  Also in the model of Szabo, Schulten and Schulten (SSS) \cite{szabo}, in which the difficult problem of the polymeric chain having many degrees of freedom is replaced with a single particle diffusing in a potential of mean force predicts $N^{3/2}$ scaling for a chain without hydrodynamic interaction. Portman \cite{portman} pointed out that SSS theory \cite{szabo} actually gives a lower bound to the chain closing time and WF theory \cite{wilemski} gives an upper bound. So a faster ring closure in SSS formalism as compared to WF theory prediction is expected. Experimentally the closure rate $\kappa$ for a $10-20$ residue polypeptides of the alanine-gylcine-glutamine trimer have been found out to be vary as $N^{-3/2}$ for large $N$ \cite{lapidusjpcb}. Although in accordance with SSS theory \cite{szabo}, this faster ring closing as compared to a flexible chain without hydrodynamic interaction might also be due to hydrodynamic interaction as found in our theoretical analysis here.

Thus just by looking at the length dependence of the ring closing time it is rather impossible to comment on the microscopic basis of ring closure. Similar scalings can arise due to completely different reasons. As a future problem it will be interesting to see what happens to the closing time for a chain with hydrodynamic interaction as different sink functions are used and also in the SSS formalism \cite{szabo}. It is expected since the SSS formalism gives a lower bound to the ring closing rate \cite{portman}, ring closing with the hydrodynamic interaction work in the SSS framework will make ring closing even faster as compared to in the WF framework.

\section{acknowledgement}

This work was supported by the Department of Science and Technology through the J. C. Bose fellowship project of  K. L. Sebastian. The author thanks K. L. Sebastian for illuminating discussions.

\bibliographystyle{apsrev}

\end{document}